\documentclass[english]{article}
\usepackage[T1]{fontenc}
\usepackage[latin9]{inputenc}
\usepackage{color}
\usepackage{mathtools}
\usepackage{url}
\usepackage{amsmath}
\usepackage{amssymb}
\usepackage{esint}

\makeatletter

\usepackage{babel}

\makeatother

\usepackage{babel}
\begin{document}
\title{\textcolor{blue}{\huge{}{}On Multi-Time Correlations in Stochastic
Mechanics}}
\author{Maaneli Derakhshani\thanks{Department of Mathematics, Rutgers University. Email: maanelid@yahoo.com
and md1485@math.rutgers.edu} \ and Guido Bacciagaluppi\thanks{Department of Mathematics and Descartes Centre, Utrecht University.
Email: g.bacciagaluppi@uu.nl}}
\maketitle

\begin{abstract}
We address a long-standing criticism of the stochastic mechanics approach to quantum theory by one of its pioneers, Edward Nelson: 
multi-time correlations in stochastic mechanics differ from those in textbook quantum theory. We elaborate upon an answer to this 
criticism by Blanchard \emph{et al.} (1986), who showed that if the (derived) wave function in stochastic mechanics is assumed to 
collapse to a delta function in a position measurement, the collapse will change the stochastic process for the particles (because the 
stochastic process depends on derivatives of the wave function), and the resulting multi-time correlations will agree with those in 
textbook quantum theory. We show that this assumption can be made rigorous through the tool of `effective collapse' familiar from 
pilot-wave theories, and we illustrate this with an example involving the double-slit experiment. We also show that in the case of 
multi-time correlations between multiple particles, effective collapse implies nonlocal influences between particles. Hence one of 
the major lingering objections to stochastic mechanics is dissolved.
\end{abstract}

\section{Introduction}

\label{intro}

Stochastic mechanics is an approach to quantum theory that aims to
recover it from an underlying stochastic process in configuration
space. As developed by Nelson (1966, 1985), and briefly described
in Section~\ref{review} for the $N$-particle case, such a recovery
starts from a time-reversible description of a diffusion process in
configuration space. By then imposing a number of (time-symmetric)
dynamical conditions on the process, most elegantly the variational
principle due to Yasue (1981), one obtains the Madelung equations
for two real functions $R$ and $S$, which can be combined into a
Schr\"{o}dinger equation for $\psi=e^{\frac{1}{\hbar}\left(R+iS\right)}$.

The resulting theory has a number of similarities with de~Broglie
and Bohm's pilot-wave theory. Indeed, particle trajectories in Nelson's
stochastic mechanics can be intuitively thought of as de~Broglie--Bohm
trajectories with a superimposed white noise (although in pilot-wave
theory $\psi$ is interpreted as a fundamental quantity). This means,
of course, that stochastic mechanics shares the same non-local features
as pilot-wave theory (one of the reasons why Nelson himself eventually
lost interest in his theory). As discussed elsewhere (Bacciagaluppi
2005), however, stochastic mechanics has a number of potential advantages
over pilot-wave theory. The main reason why it is not as popular as
de~Broglie--Bohm theory (or other fundamental approaches to quantum
theory such as spontaneous collapse or many-worlds), we surmise, is
that it is perceived as suffering from two main technical problems
not shared by other better-known approaches.

These are:

(a) The problem, raised by Wallstrom (1994), that the Madelung equations
are only equivalent to the Schr\"{o}dinger equation if one imposes a seemingly
\emph{ad hoc} quantisation condition on their solutions.

(b) The problem, raised by Nelson himself (1986), that the theory
appears to fail to recover the multi-time correlations predicted by
quantum theory.

For different reasons, however, neither of these problems are decisive.
Wallstrom's problem proved recalcitrant for many years, but after
a number of arguably only partially convincing attempts,\footnote{The most interesting probably being Schmelzer (2011).}
it has now found what appears to us to be a natural resolution, based
on combining Nelson's strategy with de Broglie's original strategy
for deriving the quantisation conditions (Derakhshani 2017). And,
as we argue in this paper, Nelson's problem of multi-time correlations
is a red herring.

The problem of multi-time correlations (which we describe in more
detail in Section~\ref{Nelson-argument}) can be expressed very intuitively
in terms of series of measurements. In standard quantum theory, in
order to calculate the correlations between the position of a particle
at multiple times, one considers a series of position measurements,
which localise the particle at the given times. Straightforward application
of the Born rule provides the desired result. In Nelson's mechanics,
the particle has a well-defined position at all times, and the stochastic
process also naturally determines the correlations in position at
multiple times. The quantum mechanical calculation and the stochastic
mechanical calculation, however, lead to different results, as spelled
out by Nelson (1986).

At the root of the problem lies the fact that in the quantum mechanical
calculation at each measurement one substitutes a new collapsed wave
function for the original unitarily evolved wave function, while the
wave function that is derived in Nelson's mechanics from the underlying
stochastic process always follows the unitary Schr\"{o}dinger evolution.
Indeed, as already pointed out by Blanchard \emph{et al.}\ (1986)
(and summarised below in Section~\ref{Blanchard-argument}), if the
corresponding substitution is made also in the stochastic mechanical
calculation, the results will coincide. The obvious question, however,
is \emph{why on earth} should one feel licensed to invoke collapse
in Nelson's theory where the wave function is not even a fundamental
object?

The answer, as we shall argue in detail in Section~\ref{effective},
lies just below the surface. Since the current velocity ${\bf v}$
in stochastic mechanics has the same form as the particle velocity
in de~Broglie--Bohm theory, many of the results and techniques developed
in the context of pilot-wave theory can be straightforwardly imported
into stochastic mechanics, even though in the latter theory the wave
function is not considered a fundamental quantity. One example, transferred
to stochastic mechanics in Bacciagaluppi (2012), is the notion of
(`sub-quantum') disequilibrium, i.e.\ the idea that frequencies in
an actual sample could differ from the theoretical probability distribution
(and consequences thereof). The pilot-wave notion we focus on in this
paper, however, is that of \emph{effective collapse}, which is the
tool developed by Bohm (1952) to describe in pilot-wave theory measurements
of quantities other than position. The key idea is that one must include
the apparatus into the pilot-wave theoretic description. Since the
trajectory of one system depends on the actual position of other systems
with which it is entangled, including information about the position
of such other systems changes the way one calculates the trajectory
of a system. And in measurement situations, including information
about the measurement result means that one can calculate trajectories
as if the wave function of the system had collapsed. As we shall discuss,
such effective collapse can be transferred also to the setting of
stochastic mechanics, and as we have suggested previously (Bacciagaluppi
2012, p.~5, fn.~8), it is the missing ingredient that allows one
to solve Nelson's problem of multi-time correlations.\footnote{The notion of 
effective collapse has been discussed in stochastic mechanics by 
Goldstein (1987) as well as by Blanchard \emph{et al.} (1991), 
but does not seem to have found wide application. Peruzzi and Rimini (1996) 
give it a unified treatment in pilot-wave theory, stochastic mechanics and 
variants thereof, and Bacciagaluppi (2003) makes essential use of it in 
discussing identical particles in pilot-wave theory and stochastic mechanics. 
Blanchard \emph{et al.} (1991, p. 162) clearly realise its relevance to the 
problem of multi-time correlations (since they reference Blanchard \emph{et al.} 
(1986)), but they do not spell out the connection explicitly. Nelson, to the best 
of our knowledge, never discussed or recognized the notion of effective collapse 
in his writings on stochastic mechanics; if he was unaware of the notion, this 
could help explain why he thought there was a fundamental discrepancy between 
stochastic mechanics and quantum mechanics for multi-time correlations.} 

We conclude that the common perception of stochastic mechanics as
suffering from major technical problems is mistaken.

\section{Review of Nelson--Yasue stochastic mechanics}

\label{review}

\
 The \emph{N}-particle version of Nelson--Yasue stochastic mechanics
(NYSM) begins with the hypothesis that 3-D space is pervaded by a
homogeneous and isotropic ether with classical stochastic fluctuations
that impart a frictionless, conservative diffusion process to a point
particle of mass $m$ and charge $e$ immersed within the ether. Accordingly,
for $N$ point particles of masses $m_{i}$ and charges $e_{i}$ immersed
in the ether, each particle will in general have its position 3-vector
$\mathbf{q}_{i}(t)$ constantly undergoing diffusive motion with drift,
as modelled by the first-order forward stochastic differential equations 

\begin{equation}
d\mathbf{q}_{i}(t)=\mathbf{b}_{i}(q(t),t)dt+d\mathbf{W}_{i}(t).\label{1}
\end{equation}
Here $q(t)=\{\mathbf{q}_{1}(t),\mathbf{q}_{2}(t),...,\mathbf{q}_{N}(t)\}$
$\in$ $\mathbb{R}^{3N}$, $\mathbf{b}_{i}(q(t),t)$ is the deterministic
mean forward drift velocity of the $i$th particle (which in general
may be a function of the positions of all the other particles, such
as in the case of particles interacting with each other gravitationally
and/or electrostatically), and $\mathbf{W}_{i}(t)$ is the Wiener
process modeling the $i$th particle's interaction with the ether
fluctuations. (The `mean' here refers to averaging over the Wiener
process, in the sense of the conditional expectation at time \emph{t}.)

The Wiener increments $d\mathbf{W}_{i}(t)$ are assumed to be Gaussian
with zero mean, independent of $d\mathbf{q}_{i}(s)$ for $s\leq t$,
and with variance 
\begin{equation}
\mathrm{E}_{t}\left[dW_{in}(t)dW_{im}(t)\right]=2\nu_{i}\delta_{nm}dt,\label{2}
\end{equation}
where $\mathrm{E}_{t}$ denotes the conditional expectation at time
\emph{t}. We then hypothesize that the magnitudes of the diffusion
coefficients $\nu_{i}$ are given by 
\begin{equation}
\nu_{i}=\frac{\hbar}{2m_{i}}.\label{3}
\end{equation}

In addition to (\ref{1}), we also have the backward stochastic differential
equations 
\begin{equation}
d\mathbf{q}_{i}(t)=\mathbf{b}_{i*}(q(t),t)dt+d\mathbf{W}_{i*}(t),\label{4}
\end{equation}
where $\mathbf{b}_{i*}(q(t),t)$ are the mean backward drift velocities,
and $d\mathbf{W}_{i*}(t))$ are the backward Wiener processes. As
in the single-particle case, the $d\mathbf{W}_{i*}(t)$ have all the
properties of $d\mathbf{W}_{i}(t)$ except that they are independent
of the $d\mathbf{q}_{i}(s)$ for $s\geq t$. With these conditions
on $d\mathbf{W}_{i}(t)$ and $d\mathbf{W}_{i*}(t)$, Eqs.\ (\ref{1})
and (\ref{4}) respectively define forward and backward Markov processes
for $N$ particles on $\mathbb{R}^{3}$ (or, mathematically equivalently,
for a single particle on $\mathbb{R}^{3N}$).

Associated to the trajectories $\mathbf{q}_{i}(t)$ is the $N$-particle
probability density $\rho(q,t)=n(q,t)/N$ where $n(q,t)$ is the number
of particles per unit volume. Corresponding to (1) and (4), then,
are the $N$-particle forward and backward Fokker-Planck equations
\begin{equation}
\frac{\partial\rho(q,t)}{\partial t}=-\sum_{i=1}^{N}\nabla_{i}\cdot\left[\mathbf{b}_{i}(q,t)\rho(q,t)\right]+\sum_{i=1}^{N}\frac{\hbar}{2m_{i}}\nabla_{i}^{2}\rho(q,t),\label{5}
\end{equation}
and 
\begin{equation}
\frac{\partial\rho(q,t)}{\partial t}=-\sum_{i=1}^{N}\nabla_{i}\cdot\left[\mathbf{b}_{i*}(q,t)\rho(q,t)\right]-\sum_{i=1}^{N}\frac{\hbar}{2m_{i}}\nabla_{i}^{2}\rho(q,t),\label{6}
\end{equation}
where we assume that the solutions $\rho(q,t)$ in each time direction
satisfy the normalization condition 
\begin{equation}
\int_{\mathbb{R}^{3N}}\rho_{0}(q)d^{3N}q=1.\label{7}
\end{equation}
Up to this point, (\ref{5}) and (\ref{6}) correspond to independent
diffusion processes in opposite time directions.\footnote{In fact, given all possible solutions to (\ref{1}), one can define
as many forward processes as there are possible initial distributions
satisfying (\ref{5}); likewise, given all possible solutions to (\ref{4}),
one can define as many backward processes as there are possible `initial'
distributions satisfying (\ref{6}). Consequently, the forward and
backward processes are both underdetermined, and neither (\ref{1})
nor (\ref{4}) has a well-defined time-reversal.} To fix the diffusion process uniquely for both time directions, we
must constrain the diffusion process to simultaneous solutions of
(\ref{5}) and (\ref{6}).

Note that the sum of (\ref{5}) and (\ref{6}) yields the $N$-particle
continuity equation 
\begin{equation}
\frac{\partial\rho({\normalcolor q},t)}{\partial t}=-\sum_{i=1}^{N}\nabla_{i}\cdot\left[\mathbf{v}_{i}(q,t)\rho(q,t)\right],\label{8}
\end{equation}
where 
\begin{equation}
\mathbf{v}_{i}(q,t)\coloneqq\frac{1}{2}\left(\mathbf{b}_{i}(q,t)+\mathbf{b}_{i*}(q,t)\right)\label{9}
\end{equation}
is the current velocity field of the $i$th particle. We shall also
require that $\mathbf{v}_{i}(q,t)$ is equal to the gradient of a
scalar potential $S(q,t)$ (since, if we allowed $\mathbf{v}_{i}(q,t)$
a non-zero curl, then the time-reversal operation would change the
orientation of the curl, thus distinguishing time directions (de la
Pe\~{n}a and Cetto 1982, Bacciagaluppi 2012)). And for particles classically
interacting with an external vector potential $\mathbf{A}_{i}^{ext}\coloneqq\mathbf{A}^{ext}(\mathbf{q}_{i},t)$,
the current velocities get modified by the usual expression 
\begin{equation}
\mathbf{v}_{i}(q,t)=\frac{\nabla_{i}S(q,t)}{m_{i}}-\frac{e_{i}}{m_{i}c}\mathbf{A}_{i}^{ext}.\label{10}
\end{equation}
So (\ref{8}) becomes 
\begin{equation}
\frac{\partial\rho({\normalcolor q},t)}{\partial t}=-\sum_{i=1}^{N}\nabla_{i}\cdot\left[\left(\mathbf{\frac{\nabla_{\mathit{i}}\mathrm{\mathit{S\mathrm{(\mathit{q},\mathit{t})}}}}{\mathit{m}_{\mathit{i}}}}-\frac{e_{i}}{m_{i}c}\mathbf{A}_{i}^{ext}\right)\rho(q,t)\right],\label{11}
\end{equation}
which is now a time-reversal invariant evolution equation for $\rho$.

The function \textit{S} is an $N$-particle velocity potential, defined
here as a field over the possible positions of the particles (hence
the dependence of $S$ on the generalized coordinates $\mathbf{q}_{i}$),
and generates different possible initial irrotational velocities for
the particles via (\ref{10}). No assumptions are made at this level
as to whether or not $S$ can be written as a sum of single-particle
velocity potentials. Rather, this will depend on the initial conditions
and constraints specified for a system of $N$ Nelsonian particles,
as well as the dynamics we obtain for $S$.

Note also that subtracting (\ref{5}) from (\ref{6}) yields the equality
on the right hand side of 
\begin{equation}
\mathbf{u}_{i}(q,t)\coloneqq\frac{1}{2}\left[\mathbf{b}_{i}(q,t)-\mathbf{b}_{i*}(q,t)\right]=\frac{\hbar}{2m_{i}}\frac{\nabla_{i}\rho(q,t)}{\rho(q,t)},\label{12}
\end{equation}
where $\mathbf{u}_{i}(q,t)$ is the osmotic velocity field of the
$i$th particle. From (\ref{10}) and (\ref{12}), we then have $\mathbf{b}_{i}=\mathbf{v}_{i}+\mathbf{u}_{i}$
and $\mathbf{b}_{i*}=\mathbf{v}_{i}-\mathbf{u}_{i}$, which when inserted
back into (\ref{5}) and (\ref{6}), respectively, returns (\ref{11}).
Thus $\rho$ is fixed as the unique, single-time, `quantum equilibrium'
distribution for the solutions of (\ref{1}) and (\ref{4}), and evolves
by (\ref{11}).

To give physical meaning to the osmotic velocities, we postulate the
presence of an external `osmotic' potential which couples to the $i$th
particle as $R(q(t),t)=\mu U(q(t),t)$ and imparts to it a momentum
$\nabla_{i}R(q,t)|_{q=q(t)}$. This momentum then gets counter-balanced
by the ether fluid's osmotic impulse pressure, $\left(\hbar/2m_{i}\right)\nabla_{i}\ln[n(q,t)]|_{q=q(t)}$,
leading to the equilibrium condition $\nabla_{i}R/m_{i}=\left(\hbar/2m_{i}\right)\nabla_{i}\rho/\rho$
(using $\rho=n/N$), hence $\rho=e^{2R/\hbar}$ for all times.

In order to formulate the second-order dynamics of the particles,
we need to generalize to $N$ particles Nelson's (1966) construction
of mean forward and backward derivatives. This generalization is straightforwardly
given by 
\begin{equation}
D\mathbf{q}_{i}(t)=\underset{_{\Delta t\rightarrow0^{+}}}{lim}\mathrm{E}_{t}\left[\frac{q_{i}(t+\Delta t)-q_{i}(t)}{\Delta t}\right],\label{13}
\end{equation}
and 
\begin{equation}
D_{*}\mathbf{q}_{i}(t)=\underset{_{\Delta t\rightarrow0^{+}}}{lim}\mathrm{E}_{t}\left[\frac{q_{i}(t)-q_{i}(t-\Delta t)}{\Delta t}\right].\label{14}
\end{equation}
By the Gaussianity of $d\mathbf{W}_{i}(t)$ and $d\mathbf{W}_{i*}(t)$,
we obtain $D\mathbf{q}_{i}(t)=\mathbf{b}_{i}(q(t),t)$ and $D_{*}\mathbf{q}_{i}(t)=\mathbf{b}_{i*}(q(t),t)$.
We note that $\mathbf{v}_{i}(q(t),t)=\frac{1}{2}\left(D+D_{*}\right)\mathbf{q}_{i}(t)$.
The computation of $D\mathbf{b}_{i}(q(t),t)$ and $D_{*}\mathbf{b}_{i}(q(t),t)$)
is given in the appendix and yields
\begin{equation}
D\mathbf{b}_{i}(q(t),t)=\left[\frac{\partial}{\partial t}+\sum_{i=1}^{N}\mathbf{b}_{i}(q(t),t)\cdot\nabla_{i}+\sum_{i=1}^{N}\frac{\hbar}{2m_{i}}\nabla_{i}^{2}\right]\mathbf{b}_{i}(q(t),t),\label{16}
\end{equation}
\begin{equation}
D_{*}\mathbf{b}_{i*}(q(t),t)=\left[\frac{\partial}{\partial t}+\sum_{i=1}^{N}\mathbf{b}_{i*}(q(t),t)\cdot\nabla_{i}-\sum_{i=1}^{N}\frac{\hbar}{2m_{i}}\nabla_{i}^{2}\right]\mathbf{b}_{i*}(q(t),t).\label{17}
\end{equation}
We now generalize the construction given by Yasue (1981) in the single-particle
case. Using (\ref{16}--\ref{17}), and assuming that particle $i$ also couples to an external electric potential, $\Phi_{i}^{ext}\coloneqq\Phi^{ext}(\mathbf{q}_{i}(t),t)$, as well as to the other particles by the Coulomb interaction potential $\Phi_{i}^{int}:=\frac{1}{2}\sum_{j=1}^{N(j\neq i)}\frac{e_{j}}{|\mathbf{q}_{i}(t)-\mathbf{q}_{j}(t)|}$, we can construct the \emph{N}-particle generalization of Yasue's ensemble-averaged,
time-symmetric mean action: 

\begin{equation}
\begin{aligned}J & =\mathrm{E}\left[\int_{t_{I}}^{t_{F}}\sum_{i=1}^{N}\left\{ \frac{1}{2}\left[\frac{1}{2}m_{i}\mathbf{b}_{i}^{2}+\frac{1}{2}m_{i}\mathbf{b}_{i*}^{2}\right]+\frac{e_{i}}{c}\mathbf{A}_{i}^{ext}\cdot\frac{1}{2}\left(D+D_{*}\right)\mathbf{q}_{i}(t)-e_{i}\left[\Phi_{i}^{ext}+\Phi_{i}^{int}\right]\right\} dt\right]\\
 & =\mathrm{E}\left[\int_{t_{I}}^{t_{F}}\sum_{i=1}^{N}\left\{ \frac{1}{2}m_{i}\mathbf{v}_{i}^{2}+\frac{1}{2}m_{i}\mathbf{u}_{i}^{2}+\frac{e_{i}}{c}\mathbf{A}_{i}^{ext}\cdot\mathbf{v}_{i}-e_{i}\left[\Phi_{i}^{ext}+\Phi_{i}^{int}\right]\right\} dt\right],
\end{aligned}
\label{18}
\end{equation}
where $\mathrm{E}\left[...\right]$ denotes the absolute expectation.\footnote{$\mathrm{E}\left[...\right]\coloneqq\int\rho(q,t)\left[...\right]d^{3N}q$. }

Upon imposing the conservative diffusion constraint through the \emph{N}-particle
generalization of Yasue's variational principle 
\begin{equation}
J=extremal,\label{19}
\end{equation}
a straightforward computation (Derakhshani 2017, Appendix 7.1) shows
that (\ref{19}) implies 
\begin{equation}
\sum_{i=1}^{N}\frac{m_{i}}{2}\left[D_{*}D+DD_{*}\right]\mathbf{q}_{i}(t)=\sum_{i=1}^{N}e_{i}\left[-\frac{1}{c}\partial_{t}\mathbf{A}_{i}^{ext}-\nabla_{i}\left(\Phi_{i}^{ext}+\Phi_{i}^{int}\right)+\frac{\mathbf{v}_{i}}{c}\times\left(\nabla_{i}\times\mathbf{A}_{i}^{ext}\right)\right]|_{q=q(t)}.\label{20}
\end{equation}
and the stochastic accelerations 
\begin{equation}
\begin{aligned}m_{i}\mathbf{a}_{i}(q(t),t) & =\frac{m_{i}}{2}\left[D_{*}D+DD_{*}\right]\mathbf{q}_{i}(t)\\
 & =\left[-\frac{e_{i}}{c}\partial_{t}\mathbf{A}_{i}^{ext}-e_{i}\nabla_{i}\left(\Phi_{i}^{ext}+\Phi_{i}^{int}\right)+\frac{e_{i}}{c}\mathbf{v}_{i}\times\left(\nabla_{i}\times\mathbf{A}_{i}^{ext}\right)\right]|_{q=q(t)},
\end{aligned}
\label{21}
\end{equation}
for $i=1,...,N$. Applying the mean derivatives in (20), using that
$\mathbf{b}_{i}=\mathbf{v}_{i}+\mathbf{u}_{i}$ and $\mathbf{b}_{i*}=\mathbf{v}_{i}-\mathbf{u}_{i}$,
and replacing $q(t)$ with $q$ in the functions on both sides, straightforward
manipulations show that (\ref{20}) turns into 
\begin{equation}
\begin{aligned} & \sum_{i=1}^{N}m_{i}\left[\partial_{t}\mathbf{v}_{i}+\mathbf{v}_{i}\cdot\nabla_{i}\mathbf{v}_{i}-\mathbf{u}_{i}\cdot\nabla_{i}\mathbf{u}_{i}-\frac{\hbar}{2m_{i}}\nabla_{i}^{2}\mathbf{u}_{i}\right]\\
 & =\sum_{i=1}^{N}\left[-\frac{e_{i}}{c}\partial_{t}\mathbf{A}_{i}^{ext}-e_{i}\nabla_{i}\left(\Phi_{i}^{ext}+\Phi_{i}^{int}\right)+\frac{e_{i}}{c}\mathbf{v}_{i}\times\left(\nabla_{i}\times\mathbf{A}_{i}^{ext}\right)\right].
\end{aligned}
\label{22}
\end{equation}
Using (\ref{10}) and (\ref{12}), integrating both sides of (\ref{22}),
and setting the arbitrary integration constants equal to zero, we
then obtain the \emph{N}-particle quantum Hamilton--Jacobi equation
\begin{equation}
\begin{aligned}-\partial_{t}S(q,t) & =\sum_{i=1}^{N}\frac{\left[\nabla_{i}S(q,t)-\frac{e_{i}}{c}\mathbf{A}_{i}^{ext}\right]^{2}}{2m_{i}}\\
 & +\sum_{i=1}^{N}e_{i}\left[\Phi_{i}^{ext}+\Phi_{i}^{int}\right]-\sum_{i=1}^{N}\frac{\hbar^{2}}{2m_{i}}\frac{\nabla_{i}^{2}\sqrt{\rho(q,t)}}{\sqrt{\rho(q,t)}},
\end{aligned}
\label{23}
\end{equation}
which describes the total energy of the possible mean trajectories
of the particles, and, upon evaluation at $q=q(t),$ the total energy
of the actual particles along their mean trajectories. So (\ref{11})
and (\ref{23}) together define the \emph{N}-particle HJM equations.

Let us now combine (\ref{11}) and (\ref{23}) into an \emph{N}-particle
Schr\"{o}dinger equation and write down the most general form of the \emph{N}-particle
wave function. To do this, we first need to impose the \emph{N}-particle
quantization condition 
\begin{equation}
\sum_{i=1}^{N}\oint_{L}\nabla_{i}S(q,t)\cdot d\mathbf{q}_{i}=nh,\label{24}
\end{equation}
(For a detailed justification of condition (\ref{24}), cf.\ Derakhshani
(2017).) Then we can combine (\ref{11}) and (\ref{23}) into 
\begin{equation}
i\hbar\frac{\partial\psi(q,t)}{\partial t}=\sum_{i=1}^{N}\left[\frac{\left[-i\hbar\nabla_{i}-\frac{e_{i}}{c}\mathbf{A}_{i}^{ext}\right]^{2}}{2m_{i}}+e_{i}\left(\Phi_{i}^{ext}+\Phi_{i}^{int}\right)\right]\psi(q,t),\label{Schreq}
\end{equation}
where $\psi=\sqrt{\rho}e^{iS/\hbar}=e^{\frac{1}{\hbar}\left(R+iS\right)}$.
Our resolution to the problem of multi-time correlations will depend
on being able to calculate Nelson trajectories from a quantum-mechanical
wave function, so it is essential for us that (\ref{24}) is imposed.

\section{Nelson's argument}

\label{Nelson-argument}

Here we describe Nelson's argument for the case of a one-dimensional
harmonic oscillator, as discussed by Blanchard \emph{et al.}\ (1986).
Although Nelson (1986) considered the case of two non-interacting
one-dimensional harmonic oscillators with the same frequency, the
one-dimensional oscillator case illustrates the same point more simply.\footnote{Nelson's example is interesting because it brings in considerations
of nonlocality. We return to these in our concluding Section~\ref{nonlocality}.} 

Suppose that the one-dimensional oscillator is in the ground state
with wave function 
\begin{equation}
\psi(x,t)=\frac{1}{(2\pi\sigma^{2})^{1/4}}e^{\left[-\frac{1}{2}\left(i\omega_{0}t+\frac{x^{2}}{2\sigma^{2}}\right)\right]},\label{26}
\end{equation}
where $\sigma=\hbar/2m\omega_{0}$. For the oscillator in the ground
state, the forward stochastic differential equation is (Nelson 1967)
\begin{equation}
\begin{split}dX(t) & =\left(\frac{1}{m}\frac{dS(x,t)}{dx}+\frac{1}{m}\frac{dR(x,t)}{dx}\right)|_{x=X(t)}dt+dW(t)\\
 & =-\omega_{0}X(t)dt+dW(t),
\end{split}
\label{27}
\end{equation}
where $S=\frac{\hbar}{2i}\ln(\frac{\psi}{\psi^{\ast}})$ and $R=\hbar\ln|\psi|$.
Solving (\ref{27}) by quadrature yields 
\begin{equation}
X(t)=e^{-\omega_{0}t}\left[X(0)+\int_{0}^{t}e^{\omega_{0}t'}dW(t')\right],\label{28}
\end{equation}
where we assume $\omega_{0}>0$ and $m>0$. The two-time correlation
function obtained using (\ref{28}) is thus 
\begin{equation}
\mathrm{E}\left[X(0)X(t)\right]=\sigma^{2}e^{-\omega_{0}t}\label{29}
\end{equation}
for $t\geq0$. 

Now we calculate the two-time correlation from the Heisenberg picture
of textbook quantum mechanics. For the oscillator Hamiltonian 
\begin{equation}
\hat{H}=-\frac{\hbar}{2m}\partial_{x}^{2}+\frac{1}{2}m\omega_{0}^{2}\hat{x}^{2}\label{30}
\end{equation}
the Heisenberg position operator at time \emph{t} is defined as 
\begin{equation}
\begin{split}\hat{x}(t) & =e^{\frac{i}{\hbar}\hat{H}t}\hat{x}(0)e^{-\frac{i}{\hbar}\hat{H}t}\\
 & =\hat{x}(0)\cos(\omega_{0}t)+\frac{\sin(\omega_{0}t)}{m\omega_{0}}\left(-i\hbar\partial_{x}\right).
\end{split}
\label{31}
\end{equation}
Hence the commutator 
\begin{equation}
\left[\hat{x}(0),\hat{x}(t)\right]=i\hbar\frac{\sin\left(\omega_{0}t\right)}{m\omega_{0}},\label{32}
\end{equation}
which vanishes for $t=n\pi/\omega_{0}$, where $n\in\mathbb{Z}$.
For these times, and assuming the ground state wave function (\ref{26}),
the two-time correlation function becomes 
\begin{equation}
\left\langle \psi\right|\hat{x}(0)\hat{x}(t)\left|\psi\right\rangle =\left(-1\right)^{n}\sigma^{2},\label{33}
\end{equation}
which clearly differs from (\ref{29}).\footnote{Blanchard \emph{et al.}\ (1986) give the general expression for the
two-time correlation function when the operators do not commute, see their
equation (34). Our equation (32) is valid for the special times when the
operators commute.} In particular, the latter vanishes as $t\rightarrow\infty$.

In Nelson's example, since the two harmonic oscillators are non-interacting
their position operators (at the same or different times) commute
and the above analysis carries over straightforwardly, i.e., one obtains
the same disagreement in results. Another worked example, involving
a particle with an initially Gaussian wave function scattering through
a potential, is given by Blanchard \emph{et al.}\ (1986).

\section{Blanchard \emph{et al.}'s resolution}

\label{Blanchard-argument}
The Nelsonian expression (\ref{28}) has a perfectly coherent meaning
as the expectation value of the product of the position of the oscillator
at $X(0)$ and $X(t)$ when the oscillator's position is \emph{not}
measured, and for this case 
the two-time correlation is indeed the one with exponential decay,
(\ref{29}). On the other hand, as Blanchard \emph{et al.}\ (1986)
note for Nelson's theory, the collapse of a subsystem's wave function
in a position measurement (in particular the first measurement at
$t=0$) changes the mean forward drift velocity in (\ref{27}) or
(\ref{1}), which in turn changes the particle's stochastic trajectory.
The reason is that the \emph{i}th drift velocity $\mathbf{b}_{i}(q(t),t)$
depends on a wave function $\psi=e^{\frac{1}{\hbar}\left(R+iS\right)}$
satisfying the Schr\"{o}dinger equation (25) via 
\begin{equation}
\mathbf{b}_{i}(q(t),t)=\frac{\hbar}{m_{i}}\left(\mathrm{Re}+\mathrm{Im}\right)\nabla_{i}\ln\psi(q,t)|_{q=q(t)},
\end{equation}
hence the drift velocity after (say) an ideal position measurement
of a Nelsonian particle will differ from before the measurement, assuming
that the wave function collapses as a consequence of the measurement. 
With this observation in hand, it is not at all surprising that (\ref{28}) for the unmeasured oscillator's position should differ from the quantum mechanical correlation (\ref{32}), the latter referring to measurements of the oscillator's position.

In the case of the oscillator, Blanchard \emph{et al.} take its collapsed
wave function at $t=0$, when the ideal position measurement is made
with result $X_{0}$, to be a delta function 

\begin{equation}
\underset{t\rightarrow0}{\mathrm{lim}}\:\psi^{X_{0}}(x,t)=\delta\left(x-X_{0}\right)
\end{equation}
which thereafter evolves via the oscillator's Hamiltonian (\ref{30})
to some excited state at finite time $t>0$. They note that since
Schr\"{o}dinger's equation with the Hamiltonian (\ref{30}) has the propagator
(Sakurai and Napolitano 1993, pp. 116--119)

\begin{equation}
K\left(x,t;x';0\right)=\sqrt{\frac{m\omega_{0}}{2\pi i\hbar\sin\left(\omega_{0}t\right)}}\exp\left[\left(\frac{im\omega_{0}}{2\hbar}\right)\times\frac{\left(x^{2}+x'^{2}\right)\cos\left(\omega_{0}t\right)-2xx'}{\sin\left(\omega_{0}t\right)}\right],
\end{equation}
with the property that $\underset{t\rightarrow0}{\mathrm{lim}}K\left(x,t;x';0\right)=\delta\left(x-x'\right)$,
the time-evolved wave function for the oscillator at $t>0$, with
initial condition (34), is just

\begin{equation}
\psi^{X_{0}}(x,t)\coloneqq\int_{-\infty}^{\infty}K(x,t;x';0)\psi^{X_{0}}(x',0)dx'=K(x,t;X_{0},0).
\end{equation}
(Our form for the propagator in (35) differs from the form given in
Blanchard \emph{et al.} (1986); however it can be readily shown, using
the definition of the propagator (Sakurai and Napolitano 1993, pp. 116--119), that the two forms are equivalent.
Our form for the propagator makes the calculation of the drift velocity
more transparent.) Using (36) in (33), we can calculate the drift
velocity for the oscillator at time \emph{t} and obtain

\begin{equation}
b^{X_{0}}\left(x,t\right)=\omega_{0}\left[\frac{x}{\tan\left(\omega_{0}t\right)}-\frac{X_{0}}{\sin\left(\omega_{0}t\right)}\right].
\end{equation}
Inserting this drift velocity in the stochastic differential equation

\begin{equation}
dX(t)=b^{X_{0}}\left(X(t),t\right)dt+dW(t)
\end{equation}
and solving by quadrature results in a periodic oscillator trajectory
for $t>0$:

\begin{equation}
X(t)=\left[\cos\left(\omega_{0}t\right)-\sin\left(\omega_{0}t\right)\cot\left(\omega_{0}s\right)\right]X_{0}+\frac{\sin\left(\omega_{0}t\right)}{\sin\left(\omega_{0}s\right)}X(s)+\sin\left(\omega_{0}t\right)\int_{s}^{t}\frac{dW(z)}{\sin\left(\omega_{0}z\right)},
\end{equation}
for $0<s\leq t$. Clearly, for $t=n\pi/\omega_{0}$, the trajectory
$X(t)$ is just the constant $\left(-1\right)^{n}X_{0}$. So the stochastic
mechanical two-time correlation $\mathrm{E}\left[X(0)X(t)\right]$,
for $t=n\pi/\omega_{0}$ and using again the probability density $\left|\psi(x,t)\right|^{2}$
to compute the absolute expectation, where $\psi(x,t)$ is the ground-state
wave function (\ref{26}) prior to the position measurement, reads\footnote{We note that this expression also coincides with the more general
expression for the two-time correlation given by Blanchard \emph{et
al.} (1986), see equation (33) therein. } 

\begin{equation}
\mathrm{E}\left[X(0)X(t)\right]=\left(-1\right)^{n}\sigma^{2},
\end{equation}
which agrees with the Heisenberg picture result (\ref{33}).

We note that the drift velocity (34) for a wave function corresponding
to a delta function is undefined. Of course, the assumption that the
wave function collapses to a delta function in an ideal position measurement
is itself an idealization. A more realistic assumption is that the
wave function collapses to a Gaussian of some narrow width $w$. For
such a function, the drift velocity exists and is well-behaved. If
we make this more realistic assumption, i.e., that $\underset{t\rightarrow0}{\mathrm{lim}}\:\psi^{X_{0}}(x,t)=g\left(x-\widetilde{x}\right)$,
where $g\left(x-\widetilde{x}\right)$ is a normalized Gaussian function
of width $w$ centered on some point $\widetilde{x}$ at $t=0$, we
need only multiply the propagator $K\left(x,t;x';0\right)$ with an
initial wave function corresponding to $g\left(x'-\widetilde{x}\right)$
and integrate over all space (that, is over $x'$) to obtain the oscillator's
wave function at $t>0$. It is easy to confirm that, for such a wave
function, the resulting drift velocity and oscillator trajectory for
$t>0$ will be periodic. It is therefore reasonable to expect that
the two-time correlation (40) will also be reproduced.

The obvious question is whether one can justify using collapsed wave
functions in stochastic mechanics. At first sight this seems incoherent:
the wave function in stochastic mechanics is a derived notion, and
one may not introduce separate postulates for its evolution. As we
have seen, the evolution of the wave function in stochastic mechanics
is given by the Schr\"{o}dinger equation (\ref{Schreq}).

On the other hand, the stochastic mechanics calculation in the previous
section does not model measurements using the Schr\"{o}dinger equation.
Indeed, the calculation assumes that the oscillator is following the
free Schr\"{o}dinger evolution throughout. Measurement is apparently taken
into account by conditioning on $X(0)$, but that means one is assuming
that a measurement simply reveals the oscillator's position without
affecting in any way its subsequent dynamics. This assumption is surely
not licensed. One needs to model the measurement explicitly and check
whether the resulting drift equals that given by the collapsed wave
function of textbook quantum mechanics.

Blanchard \emph{et al.}\ (1986) make one possible suggestion in this
regard. Precisely since the wave function in stochastic mechanics
is a derived notion, if as a consequence of the measurement one updates
the distribution $\rho$, this will lead to a different wave function
also in stochastic mechanics. Such `non-equilibrium' constraints on
the distribution of the process can indeed be modelled using a new
wave function, but require introducing also a non-linearity in the
resulting Schr\"{o}dinger equation (Bacciagaluppi 2012).

In the next section, we shall provide a justification for the use
of effectively collapsed wave functions fashioned after that used
in de Broglie and Bohm's pilot-wave theory, explicitly modelling measurements
by inclusion of appropriate degrees of freedom of the measuring apparatus.

\section{Effective collapse resolution }

\label{effective} How does pilot-wave theory treat measurements and
recover the appearance of collapse, even though there is no `true'
collapse in the theory?

We shall start by treating a very simple case, that of performing
a position measurement by letting a particle go through a slit. Assume
the particle, initially guided by a wave $\psi(t)$, goes through
the slit at time $t=0$. At times larger than 0, the pilot wave has
developed one `transmitted' and one `reflected' component: 
\begin{equation}
\psi(t)=\psi_{T}(t)+\psi_{R}(t).\label{example1}
\end{equation}
It is important to note that these two components do not overlap.
As a consequence, if the particle indeed goes through at $t=0$, for
times $t>0$ the only component of the pilot wave that is relevant
for the motion of the particle is the transmitted component. By the
same token, if the particle fails to go through, the only component
of the pilot wave that is relevant for the future motion of the particle
is the reflected component. For all effects and purposes (which in
pilot-wave theory means: for the purpose of guiding the motion of
the particle), the passage or failed passage through the slit has
collapsed the wave.

If we translate this to the framework of stochastic mechanics, we
see that through the simple inclusion of the external potential, conditioning
on ${\bf x}(0)$ in effect picks out a collapsed wave function. If
we compare the predictions of stochastic mechanics with those of quantum
mechanics, the predictions of the two theories are identical. But
in this simple case it is also unsurprising, because the quantum mechanical
predictions for later position measurements are insensitive to whether
or not we actually detect the particle at the slit and collapse the
wave function. Thus, it is actually immaterial whether stochastic
mechanics in this case is able to recover the notion of collapse.

This simple model, simply using an external potential, is essentially
how de~Broglie (1928) could treat diffraction phenomena. But it fails
already for slightly more complex measurements.

Imagine a double-slit experiment. Simply including the external potential
covers the case in which both slits are open. But now it makes all
the difference whether or not we detect \emph{which} open slit the
particle goes through. In order to model this in pilot-wave theory
(and by extension in stochastic mechanics), we need to include also
\emph{apparatus degrees of freedom}.

Let us do so, and let the particle interact with an `apparatus' with initial wave function $\phi({\bf y},t)$ for
measuring the particle's position to within the aperture of the U(pper)
or L(ower) slit. We schematically model the interaction as: 
\begin{equation}
\psi({\bf x},t)\phi({\bf y},t)\;\rightarrow\;a\,\psi_{U}({\bf x},t)\phi_{U}({\bf y},t)+b\,\psi_{L}({\bf x},t)\phi_{L}({\bf y},t).\label{example2}
\end{equation}
Note again that the two components of the wave do not overlap. So,
analogously as above, if the particle goes through the upper or lower
slit its motion is guided, respectively, by the effectively collapsed
wave functions $\psi_{U}({\bf x},t)$ or $\psi_{L}({\bf x},t)$. This
time, however, this is crucially due to the explicit inclusion of the
apparatus degrees of freedom: $\psi_{U}({\bf x},t)$ and $\psi_{L}({\bf x},t)$
will spread out from the slits and overlap, but the macroscopically
different $\phi_{U}({\bf y},t)$ and $\phi_{L}({\bf y},t)$ do not
(or negligibly so), thus ensuring the non-overlap of the two components
in (\ref{example2}).

Again, translating to the case of stochastic mechanics, conditioning
on ${\bf x}(0)$ effectively picks out the collapsed wave function
$\psi_{U}({\bf x},t)$ or $\psi_{L}({\bf x},t)$, and the predictions
of stochastic mechanics coincide with those of standard quantum mechanics
with collapse. This example generalises to arbitrary sequences of
position measurements, thereby solving Nelson's problem of multi-time
correlations: whenever position is actually measured in such a way
that the measurement result is encoded in the degrees of freedom of
some macroscopic object, the multi-time correlations calculated in
stochastic mechanics will coincide with those calculated in standard
quantum mechanics.

For completeness, let us consider also measurements of observables
other than position. Take for instance a measurement not of the position
but of the component of the particle's momentum parallel to the slits
(maybe using a suspended double slit). This is modelled as: 
\begin{equation}
\psi({\bf x},t)\phi({\bf y},t)\;\rightarrow\;c\,\psi_{\uparrow}({\bf x},t)\phi_{\uparrow}({\bf y},t)+d\,\psi_{\downarrow}({\bf x},t)\phi_{\downarrow}({\bf y},t).\label{example3}
\end{equation}
As in the case of (\ref{example2}), even though of course $\psi_{\uparrow}({\bf x},t)$
and $\psi_{\downarrow}({\bf x},t)$ overlap, the two components of
(\ref{example3}) do not overlap because $\phi_{\uparrow}({\bf y},t)$
and $\phi_{\downarrow}({\bf y},t)$ do not. Therefore, after the measurement
the particle will be guided by either $\psi_{\uparrow}({\bf x},t)$
or $\psi_{\downarrow}({\bf x},t)$, depending on the result of the
measurement as encoded in $\phi_{\uparrow}({\bf y},t)$ and $\phi_{\downarrow}({\bf y},t)$,
respectively.

In the case of stochastic mechanics, conditioning on ${\bf x}(0)$
no longer picks out a collapsed wave function, but the measurement
was not a measurement of position, so ${\bf x}(0)$ does not encode
the result of the measurement. It is ${\bf y}(0)$ that does, and
conditioning on ${\bf y}(0)$ accordingly picks out the collapsed
wave function $\psi_{\uparrow}({\bf x},t)$ or $\psi_{\downarrow}({\bf x},t)$.

The treatment of measurements and of `effective collapse' we have
just sketched was Bohm's (1952) crucial contribution to pilot-wave
theory. It allows the theory to recover the predictions of quantum
mechanics in all cases of measurement, as long as the measurement
results are encoded in some macroscopic positions (whether the positions
of pointers of or ink drops on a printout).

In stochastic mechanics, it explains how the drift velocity ${\bf b}$
after a measurement is the one corresponding to the usual collapsed
wave function of standard mechanics. The problem of multi-time correlations
disappears.

\section{Conclusion}

\label{nonlocality} Of the two traditional problems besetting stochastic
mechanics, Wallstrom's (1994) of the apparent \emph{ad hoc} character
of the quantization of the wave function and Nelson's (1986) of multi-time
correlations, Nelson himself considered the latter to be most pressing.\footnote{Personal communication to one of us (GB) at a conference in Vienna
in November 2011.} We believe both of these problems have been addressed in a satisfactory
way. In particular we have pointed out in this paper how the explicit
inclusion of measurements in stochastic mechanics resolves the problem
of multi-time correlations. Nevertheless, we presume Nelson would
not have been entirely happy, because the resolution of this problem
comes at the price of nonlocality.

In both Nelson (1986) and Nelson (2005) the problem of multi-time
correlations is presented as a Bell-type dilemma between reproducing
the quantum mechanical predictions and locality. Instead of taking
a single one-dimensional oscillator, as we did in Section~\ref{Nelson-argument},
Nelson takes two non-interacting one-dimensional oscillators of the
same frequency $\omega_{0}$ that are entangled in such a way that
at $t=0$ their positions are, say, perfectly correlated. The two
oscillators may, however, be at arbitrary separation along some other
dimension.

Since $\hat{x}_{2}(t)$ is periodic, so is the quantum mechanical
correlation function $\left\langle \psi_{0}\right|\hat{x}_{1}(0)\hat{x}_{2}(t)\left|\psi_{0}\right\rangle $.
But as with (\ref{29}) the Nelsonian correlation function $\mathrm{E}\left[X_{1}(0)X_{2}(t)\right]$
will have an exponential decay in time. Thus, in general,
\begin{equation}
\mathrm{E}[X_{1}(0)X_{2}(t)]\neq\left\langle \psi_{0}\right|\hat{x}_{1}(0)\hat{x}_{2}(t)\left|\psi_{0}\right\rangle .\label{nonlocal2}
\end{equation}
And the only way of getting equality is if the measurement on the
first oscillator at $t=0$ somehow nonlocally affects the later trajectory
of the second oscillator in some appropriate way.

As we have seen, including the measurement apparatus in the stochastic
mechanical description automatically yields the same predictions as
the standard textbook collapse. In the case of two entangled systems,
it will reproduce the collapse at a distance (as is well-known from
pilot-wave theory). Thus, a measurement on one oscillator does have
an effect on the drift velocity of the other oscillator. We note,
incidentally, that the same mechanism is what enables stochastic mechanics
to violate the Bell inequalities.\footnote{This also addresses the objection to stochastic mechanics raised by Kiukas and Werner (2010), 
who discuss violations of the Bell inequalities based on measurements of position at different times and suggest that the existence of a 
joint multi-time distributions in stochastic mechanics would prevent it from violating these Bell inequalities.} In conclusion, stochastic mechanics
is nonlocal and the dilemma is resolved in the way Nelson would have liked least.

\section*{References}
\noindent G. Bacciagaluppi (2003), `Derivation of the symmetry postulates for 
identical particles from pilot-wave theories', \url{https:\\arxiv.org/pdf/quant-ph/0302099.pdf}    

\

\noindent G. Bacciagaluppi (2005), in \emph{Endophysics, Time, Quantum and
the Subjective} (Singapore: World Scientific), pp.~367--388. Revised version: \url{http://philsci-archive.pitt.edu/8853/}

\

\noindent  G. Bacciagaluppi (2012), \emph{J. Phys. Conf. Ser.} \textbf{361} 012017, 1--12

\

\noindent P. Blanchard, M. Cini and M. Serva (1991), in \emph{BiBoS--463/91} (Bielefeld: BiBoS), pp.~149--171  

\

\noindent P. Blanchard, S. Golin and M. Serva (1986), \emph{Phys. Rev. D} \textbf{34}(12),
3732--3738

\

\noindent 
 D. Bohm (1952), \emph{Phys. Rev.} \textbf{85}(2),  166--179 and 180--193

\ 

\noindent 
L. de Broglie (1928), in \emph{Electrons et photons} (Paris: Gauthier-Villars), pp.~105--141 

\

\noindent 
 M. Derakhshani (2017), PhD Thesis (Utrecht), \url{https://arxiv.org/abs/1804.01394}

\

\noindent S. Goldstein (1987), \emph{J. Stat. Phys.} \textbf{47}(5/6), 645--667  

\ 

\noindent J. Kiukas and R. F. Werner (2010),  \emph{J. Math. Phys.} \textbf{51}(7) 072105, 1--16

\

\noindent 
 E. Nelson (1966), \emph{Phys. Rev.} \textbf{150}(4), 1079--1085

\ 

\noindent 
 E. Nelson (1967), \emph{Dynamical Theories of Brownian Motion} (Princeton: PUP)

\

\noindent 
 E. Nelson (1985), \emph{Quantum Fluctuations} (Princeton: PUP)

\

\noindent 
 E. Nelson (1986), `Field theory and the future of stochastic mechanics',
\url{https://doi.org/10.1007/3540171665_87}

\ 

\noindent 
 E. Nelson (2005), `The mystery of stochastic mechanics', \url{https://web.math.princeton.edu/~nelson/papers/talk.pdf}

\ 

\noindent 
 L.~de la Pe\~{n}a and A. M.~Cetto (1982), 
{\em Found.\ Phys.\/} \textbf{12}, 1017--1037

\

\noindent G. Peruzzi and A. Rimini, \emph{Found. Phys. Lett.} \textbf{9}(6), 505--519  

\

\noindent 
J. J. Sakurai and J. Napolitano (1993), \emph{Modern Quantum Mechanics}, revised edition (Boston: Addison-Wesley)

\ 

\noindent 
 I. Schmelzer (2011), `An answer to the Wallstrom objection against
Nelsonian stochastics', \url{https://arxiv.org/abs/1101.5774v3}

\

\noindent 
 T. Wallstrom (1994), \emph{Phys. Rev. A} \textbf{49}(3), 1613--1617

\

\noindent 
 K. Yasue (1981), \emph{J. Math. Phys.} \textbf{22}(5), 1010--1020

\appendix
\noindent 

\section{Computation of forward and backward stochastic derivatives}

To compute $D\mathbf{b}_{i}(q(t),t)$ (or $D_{*}\mathbf{b}_{i}(q(t),t)$)
expand $\mathbf{b}_{i}$ in a Taylor series up to terms
of order two in $d\mathbf{q}_{i}(t)$: 
\begin{equation}
\begin{aligned}d\mathbf{b}_{i}(q(t),t) & =\frac{\partial\mathbf{b}_{i}(q(t),t)}{\partial t}dt+\sum_{i=1}^{N}d\mathbf{q}_{i}(t)\cdot\nabla_{i}\mathbf{b}_{i}(q,t)|_{q=q(t)}\\
 & +\sum_{i=1}^{N}\frac{1}{2}\underset{n,m}{\sum}d\mathbf{\mathit{q}}_{in}(t)d\mathbf{\mathit{q}}_{im}(t)\frac{\partial^{2}\mathbf{b}_{i}(q,t)}{\partial\mathbf{\mathit{q}}_{in}\partial\mathit{q}_{im}}|_{q=q(t)}+\ldots.
\end{aligned}
\label{15}
\end{equation}
From (\ref{1}), we can replace $dq_{i}(t)$ by $dW_{i}(t)$ in the
last term, and when taking the conditional expectation at time \emph{t}
in (\ref{13}), we can replace $d\mathbf{q}_{i}(t)\cdot\nabla_{i}\mathbf{b}_{i}|_{q=q(t)}$
by $\mathbf{b}_{i}(\mathbf{q}(t),t)\cdot\nabla_{i}\mathbf{b}_{i}|_{q=q(t)}$
since $d\mathbf{W}_{i}(t)$ is independent of $\mathbf{q}_{i}(t)$
and has mean 0. From (\ref{2}), we then obtain 
\begin{equation}
D\mathbf{b}_{i}(q(t),t)=\left[\frac{\partial}{\partial t}+\sum_{i=1}^{N}\mathbf{b}_{i}(q(t),t)\cdot\nabla_{i}+\sum_{i=1}^{N}\frac{\hbar}{2m_{i}}\nabla_{i}^{2}\right]\mathbf{b}_{i}(q(t),t),\label{16-1}
\end{equation}
and likewise 
\begin{equation}
D_{*}\mathbf{b}_{i*}(q(t),t)=\left[\frac{\partial}{\partial t}+\sum_{i=1}^{N}\mathbf{b}_{i*}(q(t),t)\cdot\nabla_{i}-\sum_{i=1}^{N}\frac{\hbar}{2m_{i}}\nabla_{i}^{2}\right]\mathbf{b}_{i*}(q(t),t).\label{17-1}
\end{equation}

\end{document}